\documentstyle[preprint,aps]{revtex} 
\begin{document} 
\title{ Bremsstrahlung by static charges outside a static black hole ?} 
\author{George E.A. Matsas}
\address{Instituto de F\'\i sica Te\'orica, 
         Universidade Estadual Paulista,\\
         Rua Pamplona 145,
         01405-900-S\~ao Paulo, S\~ao Paulo,
         Brazil}
\author{Daniel Sudarsky}
\address{Instituto de Ciencias Nucleares,
         Universidad Nacional Aut\'onoma de M\'exico,\\
        A. Postal 70/543, M\'exico D.F. 04510,
       Mexico}
\author{Atsushi Higuchi}
\address{Institut f\"ur Theoretische Physik, Universit\"at Bern,\\
          Sidlerstrasse 5, CH--3012 Bern, Switzerland}

\def\baselinestretch{1.5}

\maketitle 
\begin{abstract}
We show that in complete analogy with the usual bremsstrahlung process,
when studied from the coaccelerated observer's point of view, a  charge
moving along the integral curves of the static Killing field in
the exterior of 
a static black hole gives rise to the emission of zero-energy photons, 
induced by the thermal bath of Hawking Radiation.

\end{abstract}  
\newpage

The relation between acceleration, radiation, 
and the equivalence 
principle has for a long time been the 
source of much confusion and discussion. 
In particular, in the case
of flat spacetime, 
the natural question that arises is  related to the  
compatibility of
the following two facts: First, an accelerated charge 
is known to radiate 
from the 
point of view of Minkowski observers, and second,
 according to the equivalence principle, the
 same charge is 
seen by comoving observers as a static charge in
a ''gravitational field".
This question has been sucessfully answered, first in
the classical context
by Rohrlich\cite{Roh} and futher elaborated by
Boulware\cite{B}, who showed 
that the presence of an horizon for the 
collection of
 comoving observers describing the spacetime as static 
 serves to  explain 
 the apparent contradictions, 
as due to the fact that the radiation zone (as described by
the Minkowski observers)
lies completely beyond the comoving observers horizon and is 
thus unobservable for them.
In the  quantum 
mechanical context, the solution to the apparent paradox 
(which now is cast in terms of photon emission
rates) has been 
solved by the authors\cite{HMS}, by recalling that, as
seen by the comoving observers
the static charge  (which has in fact constant proper 
acceleration)
is immersed now in a  Fulling-Davies-Unruh thermal bath 
\cite{D,U,F,UW}  that describes,
according to such observers, the Minkowski vacuum state of
the Maxwell field,
and that the interaction of the static charge with this bath 
of particles results 
in the stimulated emission
of zero energy Rindler photons that completely account for 
the ordinary
QED bremsstrahlung.

The purpose of this paper is to note that, in complete
analogy to the result
obtained in the case of the static charge in Rindler
spacetime, which 
interacts with the 
thermal bath representing the Minkowski vacuum, the 
analysis of  a static
charge in a static black hole spacetime, which interacts with a 
thermal bath representing the Hartle-Hawking
vacuum, yields a finite and nonzero response rate that 
is completely analogous to the result known to correspond to  
bremsstrahlung.

The procedure used to obtain the rate of emission and
absorption into and
from the thermal bath that was sucessfully employed in \cite{HMS} 
is the following: First, we note that the
introduction of a regulator  is required in order to make
sense of the expression
of the form 
$0 \times \infty$ that arises from the simultaneous 
consideration of the
rate of emission into the Rindler vacuum of 
''zero energy photons", and the
emission-stimulating effect of ''the number of  
zero energy photons 
present in the thermal bath". The regulating procedure  
consisted in
 introducing a fictitious oscillation in the value of the 
 charge, with
frequency $E$,
 and then to take the limit $E\to 0$ 
 after the corresponding rate was evaluated.
In this  analysis, 
we will proceed in precisely the same fashion, with the 
added complication 
 that, due to the fact that we do not know the explicit form of the
 mode functions
in the  Schwarzschild metric, we need to introduce 
a simulated potential
 that mimics the main features of the true effective potential,
  and that is simple enough that the modes can be found explicitly.

We will concentrate  for simplicity in the case of  a  scalar
field $\Phi $ 
 that interacts with a source $j$ and is described by the action:
\begin{equation}
 {\cal S}_0= \int_M d^4x \sqrt{-g}( \nabla^{\mu}\Phi \nabla_{\mu} \Phi
 +j \Phi)
\label{action1}
\end{equation}
with the background   spacetime corresponding to a  Schwarzschild 
black hole with mass $M$ and horizon at $r_H = 2M$. 
Thus  we write the metric :
\begin{equation}
ds^2 = (1-2M/r)dt^2 - (1-2M/r)^{-1}dr^2 -                                       
r^2 ( d\theta^2  +sin^2 \theta d\varphi^2  ) .                                       
\label{metric}
\end{equation}

We proceed to quantize the free
scalar field in a canonical way using for this the notion of
positive energy provided by the Killing field 
$ \partial / \partial t $. We will thus consider
the field only in the exterior
of the black hole.
The scalar field  satisfies the equation of motion in the
corresponding 
spacetime metric:

\begin{equation}
\Box \Phi_0 (x) = 0 .
\label{KG}
\end{equation}
The solution of (\ref{KG}) can be written in terms of the 
positive frequency modes 
\begin{equation}
u_{\omega l m} = {\psi_{\omega l}(r) 
\over r} Y_{lm} (\theta, \phi) 
e^{ -i\omega t} . 
\label{SKG}
\end{equation}
 Here
$\psi_{\omega l}(r)$ is the 
solution of the ordinary equation
\begin{equation}
\frac{d}{dr} 
\left[ f(r) \frac{d\psi_{\omega l}(r)}{dr} \right] - s(r)  
\psi_{\omega l}(r) - l(l+1) 
\frac{\psi_{\omega l}(r)}{r^2} = 0 ,
\label{RPWE}
\end{equation}
where $ f(r)=(1-2M/r)$ and $s(r) = -f^{-1}(r) 
\omega^2 +  2M/ r^{3} $.

For each  pair of values $\omega ,l $ there are actually two 
independent solutions of (\ref{RPWE}). We choose to use as mode
I the mode
that is purely incoming from the past horizon ${\cal H^-}$,
 and as mode II the mode that is 
purely incoming from past null infinity ${\cal J^-}$. These
two modes are then 
automatically orthogonal to
each other (with respect to the natural Klein Gordon 
inner product).

Thus, it is possible to expand the scalar field in terms of 
positive and negative energy modes as 
\begin{equation} 
\Phi_{\omega l m} (t,r,\theta, \phi) =
\sum_{l, m, \alpha}
\int_0^{+\infty} d\omega 
 (a_{\omega l m \alpha} u_{\omega l m \alpha} + H.c.) ,
\label{QF}
\end{equation}
where  
$a_{\omega l m \alpha}$ and $a^\dagger_{\omega l m \alpha}$ 
are annihilation and creation 
operators of particles with quantum numbers $\omega, l,  m$ and  
$\alpha =I, II$. 
They satisfy the usual commutation
relations \begin{equation}
[a_{\omega l m \alpha} , a^\dagger_{\omega' l' m'\alpha'}] = 
\delta (\omega - \omega')
\delta_{l l'}
\delta_{m m'} \delta_{\alpha \alpha'}.
\label{CR}
\end{equation}
In order for (\ref{CR}) to follow from the canonical 
commutation relations
for the field and its conjugate momentum, 
it is necessary  that 
the normal modes (\ref{SKG}) are Klein Gordon 
orthonormalized \cite{BD}.
This normalization condition  corresponds to:
\begin{equation}
(\omega + \omega')
\int_{r_H}^{+\infty} dr f^{-1}(r) 
\psi_{\omega l} (r) \psi^*_{\omega' l} (r)  =
      \delta(\omega - \omega') .  
\label{KGN2}
\end{equation}

Making use of the mode equation  (\ref{RPWE}), we rewrite the 
condition
(\ref{KGN2}) in the form
\begin{equation}
\lim_{L \to \infty} \left\{  
\left[ 
       -\psi_{\omega l} \frac{d \psi^*_{\omega' l}}{dr} 
       +\psi^*_{\omega' l} \frac{d \psi_{\omega l}}{dr}
\right]
       \frac{f(r)}{(\omega' - \omega) } \right\}_{r_H}^L
                 = 
\delta(\omega - \omega') .
\label{FKGN}
\end{equation}

Now, we consider the interaction of the field with a
 charge $q$, that is at rest with respect to static observers,
 i.e, it is following an orbit of the timelike 
 Killing field $\partial_t$ .
This  is described, 
in the above coordinates, by the density
\begin{equation}
j(x) = q \delta(r-r_0) 
\delta(\theta -\theta_0) 
\delta(\phi -\phi_0)/ \sqrt h  ,
\label{SC}
\end{equation}
where $h = - det(h_{\mu \nu}) $, 
and $h_{\mu \nu}$ is the spatial metric induced 
over the  equal time hypersurface $\Sigma_t$. 
With this definition, we obtain 
\begin{equation}
\int_{\Sigma_t} j = q ,
\end{equation}
for any $\Sigma_t$, where the natural volume element 
over $\Sigma_t$ is understood.

Our aim is to evaluate the particle emission 
and absorption rates 
to and from the thermal bath in which the charge is  immersed.  
Since the charge is static, it is clear that the 
spontaneous emission rate vanishes.
However, this does not imply that the induced 
emission and absorption rates must also vanish. 
This is because
these rates will depend on the number of 
particles  present in the bath, and which 
interact effectively with the source. 
In our case, the relevant modes are the zero--energy 
modes, because the static 
current (\ref{SC}) cannot interact with any of the other modes.
Since the number of zero--energy modes per unit volume in a 
thermal bath diverges, the induced emission 
and absorption rates are indefinite. 
Here, as we mentioned before, we will use the 
same ``regularization procedure'' used in \cite{HMS} 
and \cite{RW}, because it has already led to physically 
and mathematically 
sound results.
The procedure consists in replacing the static 
current (\ref{SC}) by an oscillating  one  

\begin{equation}
j(x) = q \sqrt 2 \cos[t  E] \delta (r-r_0) 
\delta (\theta - \theta_0) \delta (\phi - \phi_0)
/\sqrt{h} ,
\label{OC}
\end{equation}

and taking the limit $E \to 0$ at the end of the calculations. 
The  $\sqrt 2$ 
factor appears because of the fact that, 
at tree level, 
the emission and absorption response rates are a  
function of $q^2$, and, by the requirement 
 that the time average  of 
the {\em square} of this current is 
equal to $q^2$, the charge interacts with the
scalar field through the interaction corresponding to the second 
term in (\ref{action1}).

Now, let us calculate the emission amplitude of a Boulware mode 
$\vert \omega l m \alpha \rangle$, when our source is in the Boulware
vacuum $\vert 0 \rangle$; i.e., the quantum state annihilated by 
$ a_{\omega l m \alpha} $.
At  tree level, we have
\begin{equation}
{\cal A}^{\mbox{\scriptsize em}}_{\omega l m \alpha} 
= 
\langle \omega l m \alpha \vert {\cal S}_I \vert 0 \rangle .
\label{EA}
\end{equation}
Therefore, we obtain
\begin{equation}
{\cal A}_{\omega l m \alpha }^{\mbox{\scriptsize em}} =
 q \sqrt{2 \pi^2 f(r_0)/r_0^2\;}\;
\psi_{\omega l}^{ *\alpha} (r_0)  
Y^*_{l m} (\theta_0, \phi_0) 
\delta(\omega - \omega_0) ,
\label{EA2}
\end{equation}
where we have defined $\omega_0 \equiv E  $.
We recall that $(r_0, \theta_0, \phi_0)$ are the spatial 
coordinates of the charge's position. We note that
that a static charge can only interact with 
zero-energy modes, since in the limit $E\to 0$, 
the amplitude is proportional to $\delta(\omega )$.

The thermal bath is characterized by a temperature 
$\beta^{-1} = {\cal K}/2\pi$, where ${\cal K}$, the
surface gravity, $ = 1/4M$ (for a Schwarzschild black hole).
Thus, theemission rate per  total proper time
$T^{\mbox{\scriptsize tot}}$ of particles
with  fixed angular momentum is, in this case,
\begin{equation}
\frac{{\cal P}^{\mbox{\scriptsize em}}_{l m }}{T^{
\mbox{\scriptsize tot}}} = \frac{1}{T^{\mbox{\scriptsize tot}}}
\int_0^{+\infty}\sum_{\alpha}  d\omega \vert 
{\cal A}_{\omega l m \alpha}^{\mbox{\scriptsize em}} \vert^2
\left[ 1 + \frac{1}{e^{\omega \beta} -1} \right] .
\label{EP}
\end{equation}
The first term inside the brackets 
corresponds to the spontaneous emission contribution,
while the second one corresponds 
to the induced emission contribution. 
Notice that, for small $\omega$, the induced 
emission dominates over the spontaneous emission.
Substituting (\ref{EA2}) in (\ref{EP}), we obtain 
\begin{equation}
\frac{{\cal P}^{\mbox{\scriptsize em}}_{l m }}{T^{
\mbox{\scriptsize tot}}} 
        =\sum_{\alpha}
q^2 \pi f^{1/2}(r_0)(1/r_0^2) \; \vert \psi_{\omega_0 l \alpha}
(r_0) \vert^2 
\vert Y_{l m} (\theta_0, \phi_0 )\vert^2 
              \left[ 1 + \frac{1}{e^{\omega_0 \beta} -1} \right] ,
\label{EP2}
\end{equation}
where we have used $T^{\mbox{\scriptsize tot}} = 
2\pi f^{1/2}(r_0)\; \delta (0)$ \cite{IZ}. 
We are interested, however in the limit $E \to 0$,
($\omega_0 \to 0$) so (\ref{EP2}), becomes
\begin{equation}
\frac{{\cal P}^{\mbox{\scriptsize em}}_{l m }}{T^{
\mbox{\scriptsize tot}}} 
        = \sum_{\alpha}
\frac{q^2 \pi f^{1/2}(r_0) \;\vert Y_{l m} (\theta_0, 
\phi_0 )\vert^2}
     {\beta r_0^2} 
     \lim_{\omega_0 \to 0} \frac{\vert \psi_{\omega_0 l \alpha}
     (r_0) \vert^2} 
                                {\omega_0}   .
\label{F1}
\end{equation}

Analogously, the absorption rate per 
total proper time of particles with 
fixed angular momentum is
\begin{equation}
\frac{{\cal P}^{\mbox{\scriptsize abs}}_{l m }}{T^{
\mbox{\scriptsize tot}}} = \frac{1}{T^{\mbox{\scriptsize tot}}}
\int_0^{+\infty}  \sum_{\alpha} d\omega 
\vert {\cal A}_{\omega l m \alpha}^{\mbox{\mbox{\scriptsize abs}}}
\vert^2
\left[ \frac{1}{e^{\omega \beta} -1} \right]  .
\label{AP}
\end{equation}
Unitarity implies that 
${\cal A}_{\omega l m \alpha}^{\mbox{\scriptsize abs}}
= {\cal A}_{\omega l m \alpha}^{\mbox{\scriptsize em}}$,
hence, the absorption rate  of zero-energy particles
by the static scalar charge is given simply by 
\begin{equation}
\frac{{\cal P}^{\mbox{\scriptsize abs}}_{l m }}{T^{
\mbox{\scriptsize tot}}} 
        = \sum_{\alpha}
\frac{q^2 \pi f^{1/2}(r_0) \;\vert Y_{l m} (\theta_0,
\phi_0 )\vert^2}
     {\beta r_0^2}
      \lim_{\omega_0 \to 0} 
      \frac{\vert \psi_{\omega_0 l \alpha} (r_0) \vert^2} 
           {\omega_0}   .
\label{F2}
\end{equation}
We conclude that 
a  static charge outside 
a black hole emits and absorbs zero-energy 
modes with identical rates given by (\ref{F1}) , 
 which we could calculate explicitly, if
we knew the exact form of $\psi_{\omega l}^{\alpha}$ .
 
In order to get  more explicit information, in particular,
to acertain whether these rates
are finite, divergent or zero, we proceed to replace the radial
equations for the modes by
an analogous equation with the effective potential replaced by a
simulated potential with similar features (See \cite{Ag} for otheruses of this method).

It will be convenient to use the
dimensionless Wheeler tortoise coordinate
$x \equiv y +ln (y-1)$, where 
$y\equiv r/2M$.
The desired calculation for the exact Schwarzschild modes 
consists then in finding the solutions 
$\phi (x)_{\tilde\omega l}^{\alpha}$
of the equation:
 
\begin{equation}
 {{d^2} \over {d x^2}}\phi (x) + 
 [\tilde\omega^2 - V_{eff}(x)]\phi (x) =0   ,
\label{KG2}
\end{equation}
where $\tilde\omega =  2 M \omega$ and where 
the effective potential is:
\begin{equation}
V_{eff}(x)= (1-1/y) (1/y^3 +l(l+1)/y^2)   ,
\label{Pot1}
\end{equation}
 and normalize them so
\begin{equation}
lim_{L\to \infty} [\phi^* (x)_{\tilde\omega' l}
{d \over {d x}}\phi (x)_{\tilde\omega l} - 
\phi (x)_{\tilde\omega l}
{d \over {d x}} \phi^* (x)_{\tilde\omega' l}Ê]|^L_{-L} 
= 2M (\tilde\omega' -\tilde\omega ) \delta (\tilde\omega'
-\tilde\omega)  .
\label{Norm}
\end{equation}

Then to recover the modes, we just  put
\begin{equation}
\psi_{\omega l}^{\alpha} (r)=
\phi (x (r))_{\tilde\omega l}^{\alpha}   ,
\end{equation}
where $ x(r)= r/2M + ln ( r/2M -1)$ and $ \omega =\tilde\omega /2M$.
The approximate calculation for  Schwarzschild modes 
consists in solving  Eq. (\ref{KG2}),  but with the 
effective potential replaced
by a simulated potential which is simpler, and yet preserves
the essential features
of the true effective potential. We will take the
following form for the 
simulated potential:

\begin{equation}
V(x) = { {l(l+1)}\over x^2} ,
\label{Pot2}
\end{equation}
for $x>1$ and $V(x) = 0$ otherwise.

 We can now write explicitly  the two  solutions 
 coresponding to $\alpha = I, II$ for each $\omega$ and $l$.

I) The mode incoming from the past horizon $H^-$ 

\begin{equation}
\phi^I_{\tilde\omega l} (x) = a^I_{\tilde\omega l}
x h_l^{(1)} (\tilde\omega x) ,
\label{Mode1+}
\end{equation}
for $x>1$
\begin{equation}
\phi^I_{\tilde\omega l} (x) = a^I_{\tilde\omega l}
(  \beta^I_{\tilde\omega l} e^{i \tilde\omega x}
+ \gamma^I_{\tilde\omega l} e^{-i \tilde\omega x})
\label{Mode1-}
\end{equation}
for $x<1$.

2) The mode incoming from past null infinity ${\cal J}^-$:

\begin{equation}
\phi^{II}_{\tilde\omega l} (x) = a^{II}_{\tilde\omega l}
 (\beta^{II}_{\tilde\omega l} x h_l^{(2)} (\tilde\omega x) 
+\gamma^{II}_{\tilde\omega l} x h_l^{(1)} (\tilde\omega x) )
\label{Mode2+}
\end{equation}
for $x>1$ and
\begin{equation}
\phi^{II}_{\tilde\omega l} (x) = a^{II}_{\tilde\omega l} 
e^{-i \tilde\omega x}
\label{Mode2-}
\end{equation}
for $x<1$.

Here $ h^{(1)}_l (x) = j_l (x) + i \eta_l(x)$ and
 $ h^{(2)}_l (x) = j_l (x) - i\eta_l(x)$ are spherical
  Bessel functions (See \cite{AS} for properties and asymptotia).

From the continuity of the mode-functions and their derivatives at
$x=1$ we find :

\begin{equation}
 \beta^{I}_{\tilde\omega l} = (1/2) e^{-i \tilde\omega } [
  (1-i/\tilde\omega) h_l^{(1)} (\tilde\omega )  -i  
  ( h_l^{(1)})' (\tilde\omega )]
\label{beta1}
\end{equation}
\begin{equation}
 \gamma^{I}_{\tilde\omega l} = (1/2) e^{i \tilde\omega } [
  (1+i/\tilde\omega) h_l^{(1)} (\tilde\omega )  +i 
  ( h_l^{(1)})' (\tilde\omega )]  .
\label{gamma1}
\end{equation}
Here the prime indicates the derivative of the function 
with respect to its
argument.

 We note that:

\begin{equation}
|\beta^{I}_{\tilde\omega l}|^2 -|\gamma^{I}_{\tilde\omega l}|^2 
= (\tilde\omega)^{-2}  .
\label{bet-gam1}
\end{equation}

For the mode II we find 
  
\begin{equation}
 \beta^{II}_{\tilde\omega l} = (\tilde\omega^2 /2i) 
 e^{-i \tilde\omega } [
( h_l^{(1)})' (\tilde\omega )+ (i+1/\tilde\omega) h_l^{(1)}
(\tilde\omega )  ]
\label{beta2}
\end{equation}

\begin{equation}
 \gamma^{II}_{\tilde\omega l} = 
 ( -\tilde\omega^2 /2i) e^{-i \tilde\omega } [
( h_l^{(2)})' (\tilde\omega )+ (i+1/\tilde\omega) h_l^{(2)} 
(\tilde\omega )  ]    ,
\label{gamma2}
\end{equation}

 and note that
\begin{equation}
 |\beta^{II}_{\tilde\omega l}|^2 -
 |\gamma^{II}_{\tilde\omega l}|^2 = (\tilde\omega)^{2}  .
\label{bet-gam2}
\end{equation}

Thus normalizing the modes according to (\ref{Norm}) we find:

\begin{equation}
 | a^I_{\tilde\omega l}|^2  ={ M\over {2\pi \tilde\omega}} 
 |\beta^{I}_{\tilde\omega l}|^{-2 } ,
\label{NormMode1}
\end{equation}
and
\begin{equation}
 | a^{II}_{\tilde\omega l}|^2  ={{ M \tilde\omega}\over {2\pi }} 
|\beta^{II}_{\tilde\omega l}|^{-2} .
\end{equation}

For $l \not= 0$ we have for small $\tilde\omega$

\begin{equation} 
|\beta^{I}_{\tilde\omega l}|^2 \approx (1/4) 
l^2 d_l^2 \tilde\omega^{ -(2l+4)} 
\label{limbet1}
\end{equation}

\begin{equation} 
|\beta^{II}_{\tilde\omega l}|^2 \approx (1/4)
l^2 d_l^2 \tilde\omega^{ -(2l)}  .
\label{limbet2}
\end{equation}

Now to compute the response rate per unit proper time  we need the 
the wave function
as for all values of $x$  in the small  $\tilde\omega$  limit. 

First  analyze mode I 

 For $x<1$ we have:
\begin{equation}
|\phi^I_{\tilde\omega l} (x) |^2= { M \over {2\pi \tilde\omega}} 
  [2- \tilde\omega^{-2} |\beta^{I}_{\tilde\omega l}|^{-2 }
  +2|\beta^{I}_{\tilde\omega l}|^{-2 }
\Re (\beta^{I}_{\tilde\omega l}
(\gamma^{I}_{\tilde\omega l})^*e^{2i\tilde\omega x})]  ,
\end{equation}
 and after some  calculation we find
\begin{equation}
|\phi^I_{\tilde\omega l} (x) |^2= {{ M\tilde\omega}\over {\pi }} 2 
  [ (1-x) + l^{-1} ]^2  + {\cal O} ( \tilde\omega^3).
\label{SqMode1+}
\end{equation}

 For $x>1$:
\begin{equation}
|\phi^I_{\tilde\omega l} (x) |^2=
{ {2 M \tilde\omega}\over {l^2\pi }} x^{-2l}
+ {\cal O} ( \tilde\omega^3).
\label{SqMode1-}
\end{equation}

Note that $|\phi^I_{\tilde\omega l} (x) |^2$ 
is proportional to $\tilde\omega $,
therefore, as can be seen
from Eq. (\ref{F1}), this mode yields a finite and
non-zero rate.

Next we  analyze mode II ;

 For $x>1$ we have:
\begin{equation}
|\phi^{II}_{\tilde\omega l} (x) |^2 \approx 
{ M\tilde\omega\over {2\pi }} x^2
|\beta^{II}_{\tilde\omega l}|^{-2 }
\times
\end{equation}
\begin{equation}
[ |\beta^{II}_{\tilde\omega l}-\gamma^{II}_{\tilde\omega l}|^2 
\eta_l(\tilde\omega x)^2
+|\beta^{II}_{\tilde\omega l}+\gamma^{II}_{\tilde\omega l}|^2
j_l(\tilde\omega x)^2
 + 4j_l(\tilde\omega x) \eta_l(\tilde\omega x)
\Im (\beta^{II}_{\tilde\omega l}(\gamma^{II}_{\tilde\omega l})^* )]  ,
\end{equation}

and after some calculation
\begin{equation}
|\phi^{II}_{\tilde\omega l} (x) |^2 
\approx {{2 M\tilde\omega}\over {\pi }}\tilde\omega^{2l}
   x^2 ( c_l^2 /l^2) [(l+1) x^{-(l+1)} + l x^l]^2  .
\label{SqMode2+}
\end{equation}

 For $x<1$:

\begin{equation}
|\phi^{II}_{\tilde\omega l} (x) |^2 \approx 
{{ M\tilde\omega}\over {2\pi }} 
  4  \tilde\omega^{2l}/(d_l^2 l^2)  .
\label{SqMode2-}
\end{equation}

Thus $|\phi^{II}_{\tilde\omega l} (x) |^2$ is proportional 
to $\tilde\omega^{2l+1} $ and
according to Eq. (\ref{F1}) this mode yields 
zero emission and absorption rates.

Finally, we write explicitly the response rate of the charge 
to zero-energy modes
in first order perturbation theory, and with the approximation 
corresponding to the
substitution of the 
true effective potential of Eq. (\ref{Pot1}) by the simulated
potential of Eq. (\ref{Pot2}).
\smallskip
{\centerline {Case I: The charge is at $x > 1$ (or $ r/2M > 1.56 $)}}
\begin{eqnarray}
P^{em}_{l m I}(x>1)/T^{tot} =
P^{abs}_{l m I}(x>1)/T^{tot} =
           {q^2 M \over {2\pi}} 
   {| Y_{l m} (\theta_0, \phi_0)|^2 \over{ l^2}} \times \nonumber\\
           {(1 - 2M/r_0)^{1/2} \over{r_0^2}} 
          [r_0/2M  + ln (r_0/2M -1 )]^{-2l} 
\end{eqnarray}
where $(r_0, \theta_0, \phi_0)$ are the coordinates of 
the charge's position.

\smallskip
{\centerline {Case II: The charge
is at $x\leq 1$ (or $ r/2M \leq 1.56 $)}}

\begin{eqnarray}
P^{em}_{l m}(x\leq 1)/T^{tot} =
P^{abs}_{l m}(x\leq 1)/T^{tot} =
           {q^2 M \over {2\pi}} 
           | Y_{l m} (\theta_0, \phi_0)|^2  \times \nonumber\\       
           {(1 - 2M/r_0)^{1/2} \over{r_0^2}} 
          [l^{-1} +1 -r_0/2M  - ln (r_0/2M -1 )]^{2} , 
\end{eqnarray}
Notice that (i) these rates are finite and nonzero, 
(ii) if $r_0 \to +\infty$ the responses vanish
(iii) if $r_0 \to 2M$  the responses also vanish because 
$\lim_{\epsilon \to 0}  \sqrt \epsilon \ln^2 \epsilon =0$.

Also, it is interesting to note that since the only mode that 
generates a response is
the mode I (coming from the past horizon $H^-$) the response is 
going to be the
same in the Hartle Hawking vacuum or the Unruh vacuum.

The interpretation of these results is, however, not as
straightforward as in the case of
the charge undergoing constant proper acceleration in Minkowski spacetime,  
because in that case 
we have  both the Killing field associated with the comoving observers, 
and   a second Killing
field
associated with the  global inertial system, so the comparison of 
the corresponding results
confirmed the interpretation of the response rate
as  ordinary bremsstrahlung. However, the lesson learned in  that
exercise strongly supports
the interpretation of the response rate in this case as 
bremsstrahlung by static charges in 
a static black hole spacetime. It will be interesting to 
investigate if this result can 
be considered as a $0^{th}$ order approximation to the back 
reaction effect 
in the case of a black hole resulting from gravitational 
collapse, i.e. considering 
these response rates as the interaction of the
Hawking radiation with the infalling 
matter considered as quasistatic.

\section{Acknowledgements}
It is a pleasure to acknowledge helpful discussions with Prof. Robert M. Wald
with one of us (D.S.). G.M. would like to acknowledge partial support from
the Conselho Nacional de Desenvolvimento Cient\'\i fico e
Tecnol\'ogico CNPq.

\end{document}